\documentclass[conference]{IEEEtran}
\IEEEoverridecommandlockouts
\usepackage[nospace,noadjust]{cite}
\usepackage{amsmath,amssymb,amsfonts}
\usepackage{algorithmic}
\usepackage{graphicx}
\usepackage{subcaption}
\usepackage{textcomp}
\usepackage{xcolor}
\usepackage{float}
\usepackage{dsfont}
\usepackage{bm}
\usepackage{xfrac}
\usepackage{upgreek}
\usepackage{arydshln}
\usepackage{enumerate}
\usepackage{graphicx}
\usepackage{threeparttable}
\usepackage{caption}
\usepackage{multirow}
\usepackage[font=small]{caption}
\usepackage{hyphenat}
\newcommand{\be}{\begin{equation}}
\newcommand{\ee}{\end{equation}}

\newcommand{\norm}[1]{ || #1 ||}
\newcommand{\mb}[1]{\mathbf{#1}}
\newcommand{\bs}[1]{\boldsymbol{#1}}

\newcommand{\virg}[1]{\textquotedblleft#1\textquotedblright}

\newcommand{\cvec}[1]{ \mathrm{vec}\left(  #1 \right) }
\newcommand{\trace}[1]{ \mathrm{tr}\left(  #1 \right) }

\newcommand{\quadre}[1]{\left[  #1 \right]  }
\newcommand{\graffe}[1]{\left\lbrace   #1 \right\rbrace   }

\def\BibTeX{{\rm B\kern-.05em{\sc i\kern-.025em b}\kern-.08em
    T\kern-.1667em\lower.7ex\hbox{E}\kern-.125emX}}
\begin{document}

\title{Reinforcement learning-based waveform optimization for MIMO multi-target detection\\ \footnotesize 
}

\author{\IEEEauthorblockN{Li Wang}
\IEEEauthorblockA{Dept. of Electronic Engineering \\
Tsinghua University, Beijing , China\\
li-wang15@mails.tsinghua.edu.cn}
\and
\IEEEauthorblockN{Stefano Fortunati, Maria S. Greco, Fulvio Gini}
\IEEEauthorblockA{Dept. of Information Engineering \\
University of Pisa, Pisa, Italy\\
\{s.fortunati,f.gini,m.greco\}@iet.unipi.it}
}

\maketitle

\begin{abstract}
A cognitive beamforming algorithm for colocated MIMO radars, based on Reinforcement Learning (RL) framework, is proposed. We analyse an RL-based optimization protocol that allows the MIMO radar, i.e. the \textit{agent}, to iteratively sense the unknown environment, i.e. the radar scene involving an unknown number of targets at unknown angular positions, and consequently, to synthesize a set of transmitted waveforms whose related beam patter is tailored on the acquired knowledge. The performance of the proposed RL-based beamforming algorithm is assessed through numerical simulations in terms of Probability of Detection ($P_D$).   
\end{abstract}


\section{Introduction}
Reinforcement learning (RL) is an area of machine learning with connections to control theory, optimization, and cognitive sciences that potentially has a wide variety of applications in Radar Signal Processing. In short, the RL framework is a machine learning technique that allows an active learner, usually called \textit{agent}, to learn through experience (see e.g. the survey \cite{RL_Survey} and references therein). Specifically, the agent learns how to choose the best \textit{action} to achieve its goal by interacting with an unknown environment, without any pre-assigned control policy. As this brief introduction suggests, the \virg{learning loop} characterizing the RL framework has strong similarity with the Cognitive Radar (CR) iterative feedback control system \cite{Hay,Hay_p}.

In this paper, by drawing and merging elements form both the RL and CR frameworks, we analyse a possible application of some RL basic tools to a classical problem in colocated Multiple-Input Multiple-Output (MIMO) radar systems: the (cognitive) waveform optimization for multi-target detection. One of the main advantages of a (colocated) MIMO configuration with respect to the classical phased array is that each antenna element of the MIMO array is allowed to transmit different probing signals \cite{Stoica_col,MIMO_Fuhr,Fried}. As a consequence of this waveform diversity, the (colocated) MIMO beam pattern can be arbitrarily shaped in order to focus the transmitted power mainly in the angular directions of the potential targets. 

Using the RL formalism, we can say that our \textit{agent} is the MIMO radar and the \textit{environment} is the radar scene involving an unknown number of targets with unknown (angular) positions embedded in an unknown disturbance. So, the RL learning loop can be successfully exploited to optimize the transmitted waveforms in order to maximize the detection capabilities of the MIMO radar acting in this unknown scenario. In this work, the statistics of the disturbance are assumed to be known, and we focus our attention on uncertainties regarding the targets. Specifically, by transmitting a first set of orthonormal waveforms, that correspond to an omnidirectional beam pattern, the radar can acquire a first partial knowledge about the number and the positions of the targets. Then, according to the acquired information, the beam pattern can be shaped towards the targets by optimizing the second set of waveforms to be transmitted. This learning process can be iterated in order to continuously improve the radar detection capability and to adapt the waveforms to possible changes in the environments, e.g. a change in the number of targets.           
   
The rest of the paper is organized as follow. Sec. \ref{Sec_signal_mod} provides the signal model of a colocated MIMO radar while in Sec. \ref{detection}, the Generalized Likelihood Ratio Test (GLRT) for the detection problem at hand is derived. Sec. \ref{LR_alg} is the core of this work and it presents an original RL-based beamformig algorithm in the presence of white Gaussian noise with possibly unknown power. The performance of the proposed method is assessed in Sec. \ref{Num} by means of numerical simulations. Finally, some conclusions are collected in Sec. \ref{con}. 

\textit{Notation}: Italics indicates scalar quantities ($a$), lower case and upper case boldface indicate column vectors ($\mb{a}$) and matrices ($\mb{A}$), respectively. The relation $\mb{A} \succeq \mb{B}$ means that $\mb{A}-\mb{B}$ is a positive semi-definite matrix. The superscripts $*$, $T$ and $H$ indicate the complex conjugate, the transpose and the Hermitian operators. $\mb{I}_{N \times N}$ is the $N \times N$ identity matrix. With $\otimes$, $\trace{\cdot}$ and $\cvec{\cdot}$, we indicate the Kronecker product, the trace and the vectorization operator. Finally, $\norm{\cdot}$ indicates the Euclidean norm.

\section{Colocated MIMO radar signal model}\label{Sec_signal_mod}
Let us consider a MIMO radar system equipped with $N_T$ transmitters and $N_R$ receivers \cite{Stoica_col}. Both the receive and transmit subarrays are uniform linear arrays with half-wavelength element separation. Let $\mb{W} = [\mb{w}_1,\ldots,\mb{w}_{N_T}]^T$ denote the transmitted signal matrix. In particular, the $q$-th row of $\mb{W}$, i.e. $\mb{w}_q \in \mathbb{C}^N$, contains $N$ discrete samples of the transmitted waveform from the $q$-th element, with $q={1,\ldots,N_T}$. Following \cite{Fried}, let us assume that the transmitted waveforms can be expressed by a weighted sum of $N_T$ independent orthonormal baseband waveforms $\bs{\Phi} = [\bs{\phi}_1,\ldots,\bs{\phi}_{N_T}]^T$, where $\bs{\phi}_q \in \mathbb{C}^N$ is the $q$-th orthonormal baseband waveform. Then the transmitted signals matrix can be expressed as $\mb{W}=\mb{C}\bs{\Phi}$, where the weighting matrix is $\mb{C} = [\mb{c}_1,\mb{c}_2,\ldots,\mb{c}_{N_T}]^T$ and $\mb{c}_q \in \mathbb{C}^{N_T}$ is the weighting vector of the $q$-th transmit element whose power is given by $\norm{\mb{c}_q}^2$. The total transmitted power is then $P_T = \sum_{q=1}^{N_T}\norm{\mb{c}_q}^2$. The beam pattern generated by the transmitted waveforms is given by $B(\theta) = \mb{a}_T^T(\theta)\mb{R}_W\mb{a}_T^*(\theta)$ \cite{MIMO_Fuhr,Fried}, where $\mb{R}_W = \mb{C}\mb{C}^H$ is the covariance matrix of the transmitted waveforms, $\mb{a}_T(\theta) = [1, e^{j\pi\sin\theta}, \ldots, e^{j\pi(N_T-1)\sin\theta}]^T$ is the transmitter array steering vector and $\theta$ is the Direction of Arrival (DOA) of the target. By considering a particular angle-range cell and a single transmitted pulse, after the matched filter, the measurement model can be expressed as \cite{Stoica_conf,Fried}:
\be
\label{sig_mod}
\mathbb{C}^{N_R \times N_T} \ni \mb{Y} = \alpha \mb{a}_R(\theta)\mb{a}_T^T(\theta)\mb{C}  + \mb{N},
\ee
where $\alpha \in \mathbb{C}$ accounts for the Radar Cross Section (RCS) of the target, the two-way path loss and the straddling losses and $\mb{a}_R(\theta) = [1, e^{j\pi\sin\theta}, \ldots, e^{j\pi(N_R-1)\sin\theta}]^T$ is the receiver array steering vector. The noise term $\mb{N}$ is a matrix whose columns are mutually independent, zero-mean, circular complex Gaussian vectors with covariance matrix equal to $\sigma^2 \mb{I}_{N_R}$. 

The model in \eqref{sig_mod} can be rewritten in a more convenient vectorial form as:
\be
\label{vec_data_model}
\mathbb{C}^{N_TN_R} \ni \mb{y} = \cvec{\mb{Y}} = \alpha \mb{h}(\theta) + \mb{n},
\ee 
where, by using the properties of the Kronecker product, the vector $\mb{h}$ has been defined as:
\be
\mb{h}(\theta) = (\mb{C}^T\mb{a}_T(\theta)) \otimes \mb{a}_R(\theta),
\ee
while the noise vector is:
\be
\mb{n} = \cvec{\mb{N}} \sim \mathcal{CN}(\mb{0},\sigma^2\mb{I}_{N_TN_R}).
\ee

\section{Target detection and GLRT}\label{detection}
Let us now assume a discrete radar field of view divided into $LG$ angle-range cells. Specifically, we suppose that the angle $\theta$ in the signal model given in \eqref{vec_data_model} is a discrete variable with values in the set $\mathcal{L} \triangleq\{l\pi/L-\pi/2|l=0,\ldots,L-1 \}$. For each discrete angle (or \textit{angle bin}) $\theta_l \in \mathcal{L}$, we have $G$ range resolution cells, each of which will be indexed with the index $g \in \mathcal{G} \triangleq \{1, \ldots, G\}$. Moreover, we assume that the radar system transmits $K$ pulses (indexed by $k=1,\ldots,K$), each of which is characterized by the signal matrix $\mb{W}$ defined in Sec. \ref{Sec_signal_mod}. Under these assumptions, our aim is to handle the following Hypothesis Testing (HT) problem:
\be
\label{HT}
\left\lbrace 
\begin{array}{cc}
	H_0: & \mb{y}^k_{l,g} = \mb{n}^k_{l,g}\\
	H_1: & \mb{y}^k_{l,g} = \alpha^k_{l,g} \mb{h}_l + \mb{n}^k_{l,g}
\end{array} .
\right. 
\ee
By considering the noise parameter $\sigma^2$ as \textit{a priori known}, the GLR statistic $\Lambda^k_{l,g}\equiv \Lambda(\mb{y}^k_{l,g})$ can be expressed as:
\be
\label{GLRT}
\Lambda^k_{l,g} \triangleq 2 \ln \frac{\sup_{\alpha^k_{l,g} \in \mathbb{C}}p_{H_1}(\mb{y}_{l,g}|\alpha^k_{l,g})}{p_{H_0}(\mb{y}^k_{l,g})} = \frac{2}{\sigma^2}\frac{|\mb{h}_l^H\mb{y}^k_{l,g}|^2}{\norm{\mb{h}_l}^2}.
\ee

From Wilks's theorem \cite{wilks,MIMO_tabrikian}, the asymptotic distribution under $H_0$ of the GLR statistic in \eqref{GLRT} is given by:
\be
\label{asy_dist_H0}
\Lambda(\mb{y}^k_{l,g}|H_0) \overset{d.}{\underset{N_TN_R \rightarrow \infty}{\sim}} \chi_2^2,
\ee 
where $\chi_2^2$ indicates the central $\chi$-squared distribution with 2 degrees of freedom (dof). Similarly, under $H_1$, we have that:
\be
\label{asy_dist_H1}
\Lambda(\mb{y}^k_{l,g}|H_1) \overset{d.}{\underset{N_TN_R \rightarrow \infty}{\sim}} \chi_2^2(\delta^k_{l,g}),
\ee 
where $\chi_2^2(\delta)$ indicates the non-central $\chi$-squared distribution with 2 dof and a non-centrality parameter $\delta^k_{l,g}=2|\alpha^k_{l,g}|^2/\sigma^2$. 

The Probability of False Alarm ($P_{FA}$) is defined as:
\be
\label{PFA}
\begin{split}
	P_{FA} &= \mathrm{Pr} \{\Lambda^k_{l,g}>\lambda|H_0 \}\\
	& \underset{N_TN_R \rightarrow \infty}{\backsimeq}\int_{\lambda}^{\infty}p_{\Lambda^k_{l,g}}(a|H_0)da \triangleq H_{\chi_2^2}(\lambda),
\end{split}
\ee
where $p_{\Lambda}(\cdot|H_0) \equiv \chi_2^2$. Consequently, given a desired value of the $P_{FA}$, say $\overline{P_{FA}}$, the threshold $\lambda$ can be set as:
\be
\label{th}
\bar{\lambda} = H^{-1}_{\chi_2^2}(\overline{P_{FA}}),
\ee
where $H^{-1}_{\chi_2^2}$ is the inverse of the function $H_{\chi_2^2}$ defined in \eqref{PFA}. It is worth noting here that the noise power $\sigma^2$ is, in general, unknown. For this reason, we have to replace its true value in \eqref{GLRT} with a consistent estimate, say $\hat{\sigma}^2$. Remarkably, if $\hat{\sigma}^2$ is a $\sqrt{N_TN_R}$-consistent estimator, the asymptotic distributions of the GLR statistic given in \eqref{asy_dist_H0} and \eqref{asy_dist_H1} remains unchanged.  

\section{An RL-based beamforming algorithm for multi-target detection} \label{LR_alg}

In this section, we provide a full description of the RL-based beamforming algorithm. The basic RL notions and tools are thereafter recalled. Therefore, their application to the specific MIMO multi-target detection problem are discussed and analysed.  

Let us introduce a (finite) Markov decision process (MDP) as presented in \cite[Ch. 14]{book_machine_learning} and \cite{RL_Survey}. The MDP is a suitable model to describe the closed loop of interactions between the agent (the MIMO radar) and the environments, i.e. the radar scenario involving multiple targets and Gaussian noise. 

Formally, an MDP model is a tuple $\{\mathcal{S}, \mathcal{A}, P, \rho \}$, where:
\begin{itemize}
\item $\mathcal{S}$ is the (finite) sample space of the set of random states,
\item $\mathcal{A}$ is a (finite) set of actions,
\item $P: \mathcal{S} \times \mathcal{A} \times \mathcal{S} \rightarrow [0,1]$ is the state transition probability,
\item $\rho: \mathcal{S} \times \mathcal{A} \rightarrow \mathbb{R}$ is the reward function.
\end{itemize}
Finally, let us introduce the \textit{policy} $\pi: \mathcal{S} \rightarrow \mathcal{A}$ as the function that determines which action has to be taken at each state. 

The learning process can be summarized as follows. At time $k$, the agent observes the state $s_k \in \mathcal{S}$, that is considered to be a random variable with values in $\mathcal{S}$. Then, according to a specific policy $\pi$, the agent decides to take action $a_k=\pi(s_k) \in \mathcal{A}$. As a consequence of the action $a_k$, the agent will observes the state $s_{k+1} = \delta(s_k,a_k) \in \mathcal{S}$ with probability $P(s_k,a_k,s_{k+1})\triangleq \mathrm{Pr}\{s_{k+1}|s_k,a_k\}$ by receiving a reward $r_{k+1}=\rho(s_k,a_k) \in \mathbb{R}$. The function $P$ depends on the environment to be sensed and it is generally unknown. At this point, the agent has to choose the next action $a_{k+1} \in \mathcal{A}$ according to the policy $\pi$, and so on. Clearly, the critical point of this learning procedure is the choice of the \textit{optimal} policy $\pi$. As any other optimality criterion, the definition of the \virg{best} policy, say $\pi_{opt}$, relies on a score function that in the RL framework is the so-called \textit{expected return} $V_{\pi}(s)$ \cite[Ch. 14]{book_machine_learning}. Specifically, given a sequence of (random) states $s_1,\ldots,s_K$, the expected return for the policy $\pi$ is defined as:
\be \label{V_fun}
\begin{split}
	V&_\pi (s)  = E\graffe{\sum\nolimits_{\tau = 0}^{K-k} \gamma^\tau \rho(s_{k + \tau},\pi(s_{k + \tau}))|s_k=s} \\
	& = \rho(s,\pi(s)) + \gamma \sum_{s' \in \mathcal{S}}
	\mathrm{Pr}\{s'|s,\pi(s)\}V_\pi (s'),
\end{split}
\ee
where $s'=\delta(s,\pi(s))$, $\gamma \in (0,1]$ is a parameter that trades of short-term against long-term reward \cite{RL_Survey}. Specifically, if $\gamma$ is close to zero, only immediate rewards are considered. Given the score function $V_\pi (s)$, the task of the learning procedure is, therefore, to figure out the optimal policy $\pi_{opt}(s) = {\mathrm{argmax}}_{\pi}V_\pi (s)$, which can get the maximum expected return $V_{opt}(s)=\max_{\pi}V_\pi ({s})$ for each possible state value $s \in \mathcal{S}$. To this end, let us introduce the optimal \textit{state-action value function} $Q:\mathcal{S} \times \mathcal{A} \rightarrow \mathbb{R}$ as \cite[Ch. 14]{book_machine_learning},\cite{RL_Survey}:
\be
\begin{split}
\label{Q_function}
Q(s,a)  \triangleq \rho(s,a) + \gamma \sum\nolimits_{s' \in \mathcal{S}}
\mathrm{Pr}\{s'|s,a\}V_{opt} (s').
\end{split}
\ee
When both $\mathcal{S}$ and $\mathcal{A}$ are finite sets, a \textit{state-action value} matrix $\mb{Q}$, whose entries are $[\mb{Q}]_{s,a}  = Q(s,a)$, can be introduced. Given all the values of the entries of $\mb{Q}$, the optimal policy at the state $s$ can be simply defined as a sort of lookup table:
\be
\label{pi_opt}
\forall s \in \mathcal{S},\; \pi_{opt}(s) = \mathrm{argmax}_{a \in \mathcal{A}}[\mb{Q}]_{s,a}.
\ee
However, as already pointed out, the conditional probability $\mathrm{Pr}\{s'|s,a\}$ in \eqref{Q_function} is generally unknown. Then, the function $Q(s,a)$, at least in the form presented in \eqref{Q_function}, cannot be used directly to find the optimal policy. Fortunately, some additional manipulation is allowed. By definition, $V_{opt} (s') = \max_{a'\in \mathcal{A}}Q(s',a')$, so that \eqref{Q_function} can be rewritten in terms of a conditional expectation as:
\be
\label{Q_E_cond}
Q(s,a) = E_{s'}\{\rho(s,a) + \gamma
	\max_{a'\in \mathcal{A}}Q(s',a')|s\},
\ee 
which does not explicitly depend on $\mathrm{Pr}\{s'|s,a\}$. Finally, we can exploit some well-known stochastic approximation algorithms to iteratively obtain all entries of the matrix $Q(s,a)$ as discussed e.g. in \cite[Ch. 14]{book_machine_learning}. The aim of the next subsections is to show how to apply this iterative learning algorithm to the MIMO beamforming problem at hand.

\subsection{The state space} \label{state_space}
The MIMO radar detection scheme consists in testing all $LG$ angle-range cells one by one using the GLR statistic $\Lambda^k_{l,g}$ introduced in \eqref{GLRT}. Starting from the values assumed by $\Lambda^k_{l,g}$ in each $(l,g)$ angle-range cell at time $k$, the state space $\mathcal{S}$ can be set up as follows. Let us first define the statistic
\be
\label{Lambda_bar}
\bar{\Lambda}^k_l \triangleq
\left\lbrace \begin{array}{cc}
	1 & \exists g\in \mathcal{G}, \;\Lambda^k_{l,g} > \bar{\lambda} \\
	0 & \mathrm{otherwise} 
\end{array}\right. .
\ee 
In words, the statistic $\bar{\Lambda}^k_l$ is equal to 1 if the decision statistic $\Lambda^k_{l,g}$ exceeds the threshold at least in one range cell, at time step $k$, for the given $l$-th angular bin. Roughly speaking $\bar{\Lambda}^k_l$ tells us that the $l$-th angular bin may, or may not, contains some targets at time $k$. Let us now define the discrete random variable $s_k$ as:
\be
\label{s_st}
s_k \triangleq \sum\nolimits_{l}^L\bar{\Lambda}^k_l,
\ee
which tells us how many angular bins may contain targets. 

The \textit{state space} $\mathcal{S} = \{0,\ldots,T_{max}\} \subset \mathbb{N}$ is then set to be equal to the \textit{sample space} of the random variable $s_k$. Note that $T_{max}$ is the maximum number of targets, in different angle bins, say $T_{max}$, that can be identified by the MIMO radar \cite{ident_MIMO}.  

\subsection{The set $\mathcal{A}$ of the actions}

An action $a \in \mathcal{A} = \{a_i|i \in \{1,2,\ldots,T_{max}\}\}$ in the MIMO beamforming problem at hand can be defined as the combination of the two radar tasks of collecting a data snapshot $\mb{y}^k_{l,g}$ defined in \eqref{vec_data_model} and optimizing the weighting matrix $\mb{C}$ in order to focus the transmitted power on the $i$ angle bins that contain potential targets. The cardinality $T_{max}=|\mathcal{A}|$ of the set of actions $\mathcal{A}$ should then be set to be equal to the maximum number of the identifiable targets $T_{max}$. 

Let us now describe in detail a typical action that the MIMO radar at hand has to perform at each time step $k$. As said before, an action $a_k \in \mathcal{A}$ consists firstly in the acquisition of the data snapshot $\mb{y}^k_{l,g}$ and then in a more involved optimization task. Specifically, the agent (the MIMO radar) firstly has to figure out the set of $i$ angle bins $\bar{\Theta}_i = \{\bar{\theta}_1,\ldots,\bar{\theta}_{i}\} \subset \mathcal{L}$ that most likely contains the targets, and then it has to optimize the beamforming procedure, i.e. it has to find the best matrix $\mb{C}$ to synthesize a beam pattern with a power distribution based on $\bar{\Theta}_i$. It is worth underlining here that the index $i$ used to define the sequence of sets $\{\bar{\Theta}_i\}_{i=1}^{T_{max}}$ is the same index that characterizes each element of the set $\mathcal{A}$. In the following, we describe these two steps more accurately.

\subsubsection{Step 1}
Let $\bs{\Lambda}^k \in \mathbb{R}^{L \times G}$ be the matrix whose entries are the values of the GLR statistic at time $k$ for each angle-range cell, i.e. $[\bs{\Lambda}^k]_{l,g} \triangleq \Lambda^k_{l,g}$. Then, let $\mb{t} \in \mathbb{R}^{L}$ be the vector whose $l$-th entry represents the maximum value of the decision statistic over the range cells for the $l$-th angular bins, i.e. $[\mb{t}]_l = \max_{g \in \mathcal{G}}\Lambda_{l,g}$. Finally, let $\mathcal{T}_{i}$ be the set of indices of the $i$ larger values of the entries of $\mb{t}$, i.e.:
\be
\mathcal{T}_{i} \triangleq \underset{l \in \{0,\ldots,L-1\}}{\textsuperscript{\textit{i}}\mathrm{argmax}} \; \mb{t}.
\ee 
Consequently, $\bar{\Theta}_i \triangleq \{\bar{\theta}_j \in \mathcal{L}| j \in \mathcal{T}_{i} \}$.

\subsubsection{Step 2}
After having obtained the set $\bar{\Theta}_i$, the weighting matrix $\mb{C}$ has to be designed in order to focus the transmitted power on the angles bins indexed in $\bar{\Theta}_i$. The aim of the resulting optimization problem is to maximize the minimum of $B(\bar{\theta}_j) = \mb{a}_T^T(\bar{\theta}_j)\mb{R}_W\mb{a}_T^*(\bar{\theta}_j)$ with $\bar{\theta}_j \in \bar{\Theta}_i$, under the energy constraint $\trace{\mb{R}_W} = P_T$, where $\mb{R}_W = \mb{C}\mb{C}^H$. This optimization problem can be expressed as:
\be
\begin{array}{l}
	\max_{\mb{C}} \min_{j \in \mathcal{T}_{i}} \graffe{\mb{a}_T^T(\bar{\theta}_j)\mb{C}\mb{C}^H\mb{a}_T^*(\bar{\theta}_j)}\\
	\mathrm{c. \; t.} \; \trace{\mb{C}\mb{C}^H} = P_T,
\end{array}
\ee 
or, equivalently, as a semi-definite program (SDP) \cite{conv_1,conv_2} as:
\be
\begin{array}[r]{l}
	\max_{\mb{R}_W} \zeta\\
	\mathrm{c. \; t.} \; \begin{subarray}
		\mb{R}_W \succeq \mb{0}, \quad \zeta \geq 0, \quad \trace{\mb{R}_W} = P_T,\\
		\mb{a}_T^T(\bar{\theta}_j)\mb{R}_s\mb{a}_T^*(\bar{\theta}_j)\geq\zeta, \forall j \in \mathcal{T}_i
	\end{subarray},
\end{array}
\ee 
where, after having obtained $\mb{R}_W$, the weighting matrix $\mb{C}$ can be derived using the algorithm in \cite{C_mat}.

\subsection{The reward}
We indicate with $r_{k+1}$ the immediate reward obtained when the action $a_k \in \mathcal{A}$   is taken in the case of the state $s_k \in \mathcal{S}$. For the MIMO radar detection problem at hand, a reasonable reward has to be related to the overall Probability of Detection ($P_D$) for all targets. Since, in each $(l,g)$ angle-range cell at time $k$, $P_D$ can be asymptotically approximated as:
\be
(P_D)^k_{l,g}\underset{N_TN_R \rightarrow \infty}{\backsimeq} \int\nolimits_{\bar{\lambda}}^{\infty} p_{\Lambda^k_{l,g}}(a|H_1)da,
\ee
where $\bar{\lambda}$ is the threshold defined in \eqref{th} and $p_{\Lambda^k_{l,g}}(\cdot|H_1)\equiv \chi_2^2(\hat{\delta}^{k}_{l,g})$ (see \eqref{asy_dist_H1}), a possible reward function may be: 
\be
\label{rew}
r_{k+1} = \rho(s_k,a_k) \triangleq \sum_{l=0}^{L-1}\sum_{g=1}^{G} \psi(\hat{\delta}^{k}_{l,g}),
\ee
where:
\be
\psi(\hat{\delta}^{k}_{l,g}) \triangleq
\left\lbrace \begin{array}{cc}
	p_{\Lambda^k_{l,g}}(\hat{\delta}^{k}_{l,g}|H_1) & \hat{\delta}^{k}_{l,g} > \bar{\lambda} \\
	0 & \mathrm{otherwise} 
\end{array}\right. ,
\ee 
\be
\hat{\delta}^{k}_{l,g} = 2|\hat{\alpha}^{k}_{l,g}|^2/\sigma^2, \quad \hat{\alpha}^{k}_{l,g},  =(\mb{h}_l^{k})^H\mb{y}^{k}_{l,g}/\norm{\mb{h}_l^{k}}, 
\ee
and $\hat{\alpha}^{k}_{l,g}$ is the Maximum Likelihood (ML) estimate of $\alpha$ in the $(l,g)$-th angle-range cell at time step $k$.

\subsection{The \textit{SARSA} $Q$-learning algorithm}
As discussed before, the crucial step of any RL-based procedure is the choice of the optimal policy $\pi_{opt}$ defined in \eqref{pi_opt}. To this end, the state-action value function $Q$ in \eqref{Q_E_cond} has to be firstly estimated. Among all the possible stochastic approximation-based algorithms available for this task, in this work we choose to apply the so-called SARSA (state-action-reward-state-action) (see e.g. \cite{Rummery} or \cite[Sec. 14.5.4]{book_machine_learning}). In brief, SARSA is an iterative procedure involving two main steps:
\begin{enumerate}
	\item Obtain a new state $s_{k+1}=\delta(s_k,a_k)$, 
	\item Choose a new action $a_{k+1}$ through an $\epsilon$-greedy algorithm according to the current value of the $Q$-function, i.e. $Q_k$. In particular, let $a_{opt} \triangleq \mathrm{argmax}_{a \in \mathcal{A}}{Q}_k(s_{k+1},a)$ be the optimal action, then
	\be
	a_{k+1}=\left\lbrace 
	\begin{array}{ll}
		a_{opt} & \mathrm{with\; prob.\;}\epsilon\\
		a \in \mathcal{A} \setminus a_{opt}&\mathrm{with\; prob.\;}1-\epsilon\\
	\end{array}\right. .
	\ee
	Note that this $\epsilon$-greedy selection of $a_{k+1}$ is required to guarantee the convergence of the SARSA algorithm (see \cite[Ch. 14]{book_machine_learning} for more details).
	\item Starting from an initial value, say $Q_0$, update the $Q$-function according to the following iteration:
	\be
	\begin{split}
		Q_{k+1} & \leftarrow \beta Q_k(s_k,a_k) +(1-\beta) \times\\
		& \times \quadre{r_{k+1} + \gamma Q_k(s_{k+1},a_{k+1})-Q_k(s_k,a_k)},
	\end{split}
	\ee
	where $r_{k+1}$ is the reward defined in \eqref{rew}, $\beta \in (0,1)$ controls the convergence speed and $\gamma$ has been already defined in \eqref{V_fun}.
\end{enumerate}

It is worth noting that the SARSA algorithm has three free parameters, i.e $\gamma$, $\beta$ and $\epsilon$, that need to be chosen heuristically.   

\section{Numerical results} \label{Num}
In this section, the performance of the proposed RL-based algorithm is assessed through Monte Carlo simulations in two different radar scenarios. In the first study case, we simulate 4 targets with different Signal-to-Noise Ratio (SNR) that maintain the same angular position for the whole observation period of $K$ time steps. In the second study case, a scenario in which the number and the positions of the targets change during the observation period is considered. For both the study cases, we consider a colocated MIMO radar system with a uniform linear transceiver array with $N_T=N_R=16$ elements. The SNR of the $t$-th simulated target is $\mathrm{SNR}_t \triangleq \sfrac{E\{|\alpha^k_t|^2\}}{\sigma^2}$ where $\alpha^k_t \sim \mathcal{CN}(0,\sigma^2_t)$. Note that $\alpha^k_t$ has different realization from time step to time step. The noise power $\sigma^2$ is chosen to be equal to 1. To save some computational time, the radar scene is restricted in an uniform angular grid of $L=22$ angle bins between $-45^\circ$ and $45^\circ$ with $G=100$ range cells. The maximum number of identifiable targets is loosely set as $T_{max}=10$. The decision threshold $\bar{\lambda}$ is chosen for a nominal $\overline{P_{FA}}=10^{-5}$. The free parameters of the SARSA algorithm are: $\beta=0.8$, $\gamma=0.1$, $\epsilon=0.5$. The number of Monte Carlo runs is $MC=1000$.

\begin{figure}[htbp]
	\centering
	\includegraphics[height=4.5cm]{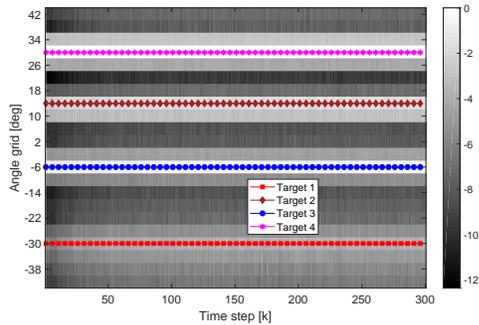}
	\caption{Normalized beam patter $D(\theta)$: study case 1.}
	\label{fig:Fig1}
\end{figure}

\textit{Study case 1:} The four \virg{fixed} targets have been generated according to the following ordered couples of angular position and SNR: Target 1 $(-30^\circ, -10\; \mathrm{dB})$; Target 2 $(14^\circ, -8\; \mathrm{dB})$, Target 3 $(-6^\circ, -6\; \mathrm{dB})$; Target 4 $(30^\circ, -4\; \mathrm{dB})$. In Fig. \ref{fig:Fig1}, we plot the averaged (over $MC$ Monte Carlo runs) normalized beam pattern defined as $D(\theta) \triangleq 10 \log_{10}(\sfrac{B(\theta)}{\max_{\theta}B(\theta)})$ for each time step $k=1,\ldots,K$. As Fig. \ref{fig:Fig1} shows, the proposed RL-based beamforming is able to exploit the information about the radar scene provided by the GLRT to focus the transmitted power on the four angle bins containing the targets. Fig. \ref{fig:Q_1} shows the absolute value of the difference between the $Q$-function estimated by the SARSA algorithm at two consecutive time steps, i.e $\xi_k \triangleq |Q_k-Q_{k-1}|$. As we can see, $\xi_k \rightarrow 0$ as $k\rightarrow\infty$ and this is an heuristic proof that the SARSA algorithm converges. Finally, in Table \ref{tab1}, we compare the Probability of Detection of the four targets, averaged over $K$ time steps, when the proposed RL-based beamforming is used against a classical omnidirectional beamformer. As expected, better detection performance is achieved by using the proposed beamforming method.

\begin{figure}[htbp]
	\centering
	\includegraphics[height=4.5cm]{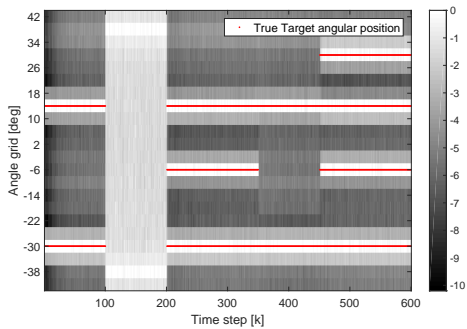}
	\caption{Normalized beam patter $D(\theta)$: study case 2.}
	\label{fig:Fig3}
\end{figure}

\begin{figure}[htbp]
	\centering
	\begin{subfigure}[b]{0.22\textwidth}
	\includegraphics[width=\textwidth]{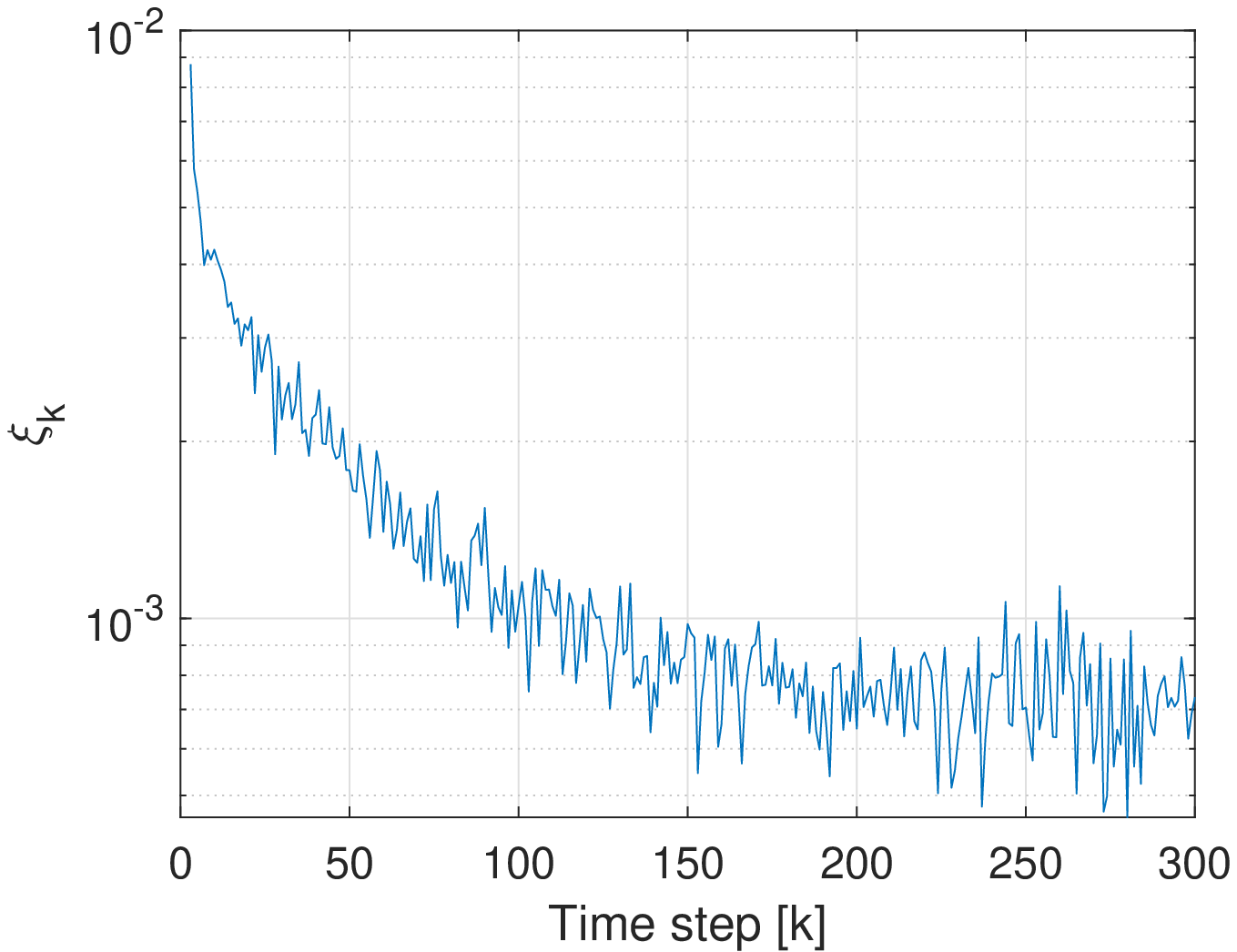}
	\caption{Study case 1}
	\label{fig:Q_1}
	\end{subfigure}
	~ 
	\begin{subfigure}[b]{0.22\textwidth}
	\includegraphics[width=\textwidth]{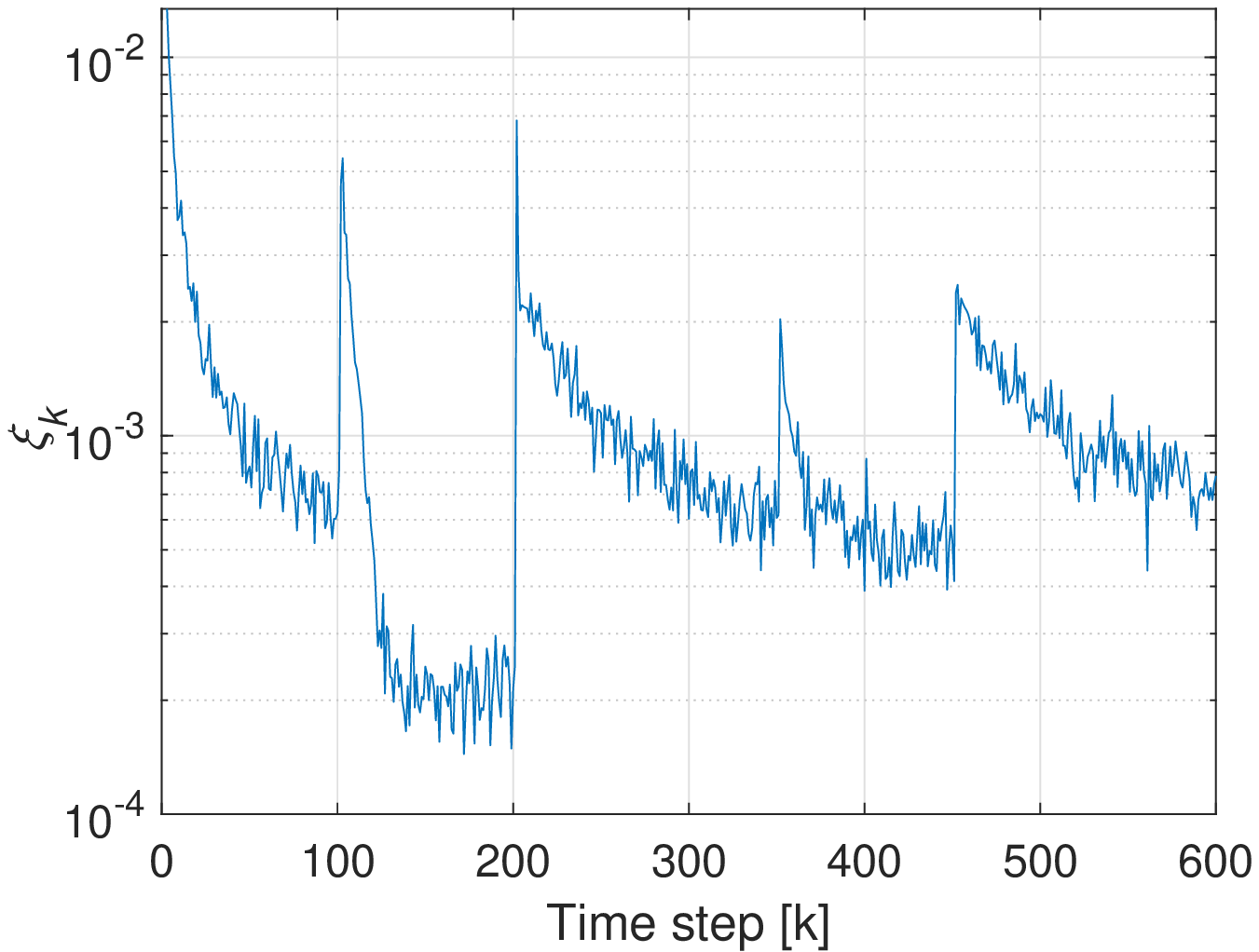}
	\caption{Study case 2}
	\label{fig:Q_2}
	\end{subfigure}
\caption{The convergence index $\xi_k\triangleq |Q_k-Q_{k-1}|$.}\label{fig:conver}
\end{figure}

\begin{table}[htbp]
	\begin{center}
\begin{tabular}{l*{4}{c}r}
	 $P_D$             & T1 & T2 & T3 & T4 \\
	\hline
	RL-based 				& 0.22 & 0.55 & 0.78 & 0.91   \\
	Omni           		 & 0.14 & 0.41 & 0.69 & 0.86  \\
\end{tabular}
\caption{Detection performance comparison: study case 1.}
\label{tab1}
\end{center}
\end{table}
\textit{Study Case 2:} In this second study case, we assess the performance of the proposed RL-based beamforming in a dynamic environment, where the number of targets and their angular positions change over the observation time. The SNR of each target is assumed to be equal to -8 dB. The radar scene is generated as follows:
\begin{itemize}
	\item $k=1 \rightarrow 100$: two targets located at $-30^\circ$ and $14^\circ$,
	\item $k=101 \rightarrow 200$: no targets are present,
	\item $k=201 \rightarrow 350$: three targets at $-30^\circ$, $-6^\circ$ and $4^\circ$,
	\item $k=351 \rightarrow 450$: two target at $-30^\circ$ and $4^\circ$,
	\item $k=451 \rightarrow 600$: four targets at $-30^\circ$, $-6^\circ$, $14^\circ$ and $29^\circ$.
\end{itemize}
As clearly shown by Fig. \ref{fig:Fig3}, the proposed RL-based beamforming algorithm is able to handle this dynamic scenario by reshaping the beam pattern according to the changes in the number of targets presented in the radar scene and in their angular locations. Finally, Fig. \ref{fig:Q_2} shows the progress of the index $\xi_k$ previously defined. It can be noted that a transition in the estimate of $Q$-function is present every time a change in the scenario occurs. However, it eventually converges after some time steps. 

\section{Conclusions} \label{con}
In this paper we have shown that some basic RL tools can be successfully applied in a dynamic MIMO radar detection problem in the presence of an environment with unknown, and variable in time, number of targets with unknown angular positions. Specifically, a RL-based waveforms optimization algorithm capable to focus the transmitted power only in the angle bins that contain possible targets has been proposed and analysed under the assumption of a priori known disturbance statistics. Future works will explore the possibility to extend the proposed algorithm by endowing it with the capability of learning the disturbance model and consequently, of optimizing the set of transmitted waveforms in order to mitigate its impact on the detection performance. 

\bibliographystyle{IEEEtran}
\bibliography{ref_Asilomar}

\end{document}